# SHG microscopic observations of polar state in Li-doped KTaO$_3$ under electric field

Hiroko Yokota*, Taeko Ohyama and Yoshiaki Uesu

Department of Physics, Waseda University

3-4-1 Okubo, Shinjuku-ku, Tokyo 169-8555, Japan

Incipient ferroelectric KTaO$_3$ with off-center Li impurity of the critical concentration of 2.8 mol% was investigated in order to clarify the dipole state under electric field. Using optical second-harmonic generation (SHG) microscope, we observed a marked history dependence of SHG intensity through zero-field cooling (ZFC), zero-field heating (ZFH), field heating after ZFC (FH/ZFC) and FH after field cooling (FH/FC). These show different paths with respect to temperature: In the ZFC/ZFH process, weak SHG was observed at low temperature, while in the FH/ZFC process, relatively high SHG appears in a limited temperature range below $T_F$ depending on the field strength, and in the FC and FH/FC processes, the SHG exhibits ferroelectric-like temperature dependence: it appears at the freezing temperature of 50K, increases with decreasing temperature and has a tendency of saturation. These experimental results strongly suggest that dipole glass state or polar nano-clusters which gradually freezes with decreasing temperature is transformed into semi-macroscopic polar state under the electric field. However at sufficiently low temperature, the freezing is so strong that the electric field cannot enlarge the polar clusters. These experimental results show that the polar nano-cluster model similar to relaxors would be more relevant in $KTaO_3$ doped with the critical concentration of Li. Further experiments on the anisotropy of SHG determine that the average symmetry of the field-induced polar phase is tetragonal 4mm or 4. The tetragonality is also confirmed by the X-ray diffraction measurement.

# I. INTRODUCTION

$KTaO_3$ is known to be one of the representatives of quantum paraelectrics.[1-4] The quantum paraelectricity is a characteristic feature exhibited by several anharmonic optical phonon systems.[5] Due to self-energy, the frequency of the lowest lying optical mode becomes strongly temperature dependent with general tendency that the frequency becomes remarkably reduced as T→0K, which is called soft optic mode. With the renormalization due to the quartic anharmonicity, there exists a finite temperature $T_0$, where the soft phonon frequency equals to zero and the high temperature phase becomes instable. Actually, however, at low temperature, the quantum fluctuations (zero point vibration) start to play a role to tend to suppress the instability. When the renormalization effect is sufficiently large, the lattice stays on cubic down to T=0K. Such a system is called quantum paraelectric. In this sense, quantum paraelectrics are generally considered incipient ferroelectrics. Therefore, only by small perturbation of external conditions such as replacement of atom, introduction of impurity, application of external field etc., the system realizes transition to dipole glass state or sometimes to ferroelectric phase at a finite temperature.

In $KTaO_3$, the introduction of Li impurity at K site by a few mol percent (expressed hereafter as KTL or KTL:$x$%Li in the case where the concentration is significantly important) induces dipole glass state as a result of off-center nature of Li in the lattice[6]. A number of studies have been directed to elucidate the structural property of the system, in particular, whether the long-range ferroelectric phase appears,[7-11] or the polar state induced by Li is localized to remain in the dipole glass state.[12-19] Now the problem seems to be settled: When the Li concentration $x$ exceeds 5 mol%, the para and ferroelectric phase boundary is well defined. In the region of 2<$x$<3, the phase boundary becomes obscured, and below the concentration, the dipole glass state is main feature of KTL.[11] As a disordered system conceptually related with spin-glass, KTL with $x$<3 is especially important and the peculiar phenomena are expected to be observed. Among them the non-ergodic behavior of the order parameter under the external field, is an essential feature of the disordered system. Although some papers described the property,[17, 19] more systematic studies are necessary to clarify it. For this purpose, we chose KTL with the critical concentration (2.8%) of Li, and observed the history dependence of the polar state developed under the electric field using the special nonlinear optical microscope (SHG microscope) which provides 2-dimensional images of optical second-harmonic(SH) waves generated in specimens.[20, 21]

SHG measurements of KTL with different concentration of Li have been so far performed, as it is quite sensitive to the appearance of a polar phase. These experiments measured the temperature dependence of SH intensity with and without electric field, for the purpose of determining the spatial correlation length from the k-dependence of the scattered SH waves,[22] or from the SHG coherence length.[19] The relaxation time and the activation energy were determined by the kinetic behavior of SH intensity after switching-off the electric field.[19, 23] The average symmetry of KTL under the electric field is also discussed[19] based on the result of polarization dependence of the fundamental wave.

This paper describes results of the history dependence of SH intensity under the electric field, i.e., zero-field cooling (ZFC), zero-field heating(ZFH), field cooling(FC) and field heating(FH) processes. Special care has been paid to erase the memory effect of specimens. The average symmetry of the polar state induced by the external electric field is determined not only from dependence of the anisotropy of fundamental waves as was previously reported but also from that of SH waves. The result is compared with the X-ray powder diffraction measurement performed down to 10K on ZFH/ZHC process.

## II. EXPERIMENTAL

A. SHG microscopic observations

Specimens used in experiments are KTL:2.8%Li, which were grown by the self-flux method with a slow-cooling technique of $Ta_2O_5$, $Li_2CO_3$ and an excess of $K_2CO_3$ as a flux.[24] Li concentration $x$ was determined by the empirical relation between $x$ and the freezing temperature $T_F$: $T_F(K) = 535(x/100)^{0.66}$.[17] $T_F$ was determined from the disappearance temperature of SHG intensity in ZFH after FC process as described later. The homogeneity of Li concentration inside specimens was checked by the distribution of $T_F$ in different illuminated places of the specimen, which is discussed later.

Specimen used in the SHG microscopic observation is a (100) plate (hereafter the axes are referred to the cubic axes), with edges parallel to the {100}, the area 9.4x5.0 mm$^2$ and the thickness 0.854 mm. Both surfaces are optically finished and two narrow rectangular electrodes apart by 3 mm are put on the top surface by Au evaporation. The orientation and dimension of the specimen are illustrated in Fig.1. A special care was paid for adjusting the illuminated place of the laser beam so as to be between two electrodes and avoid an occurrence of photocurrent, as KTL exhibits marked photocurrent effect in low temperature region.[25-31]

The optical system of SHG microscope is illustrated in Fig.2. Pulsed waves from Q switched Nd$^{3+}$:YAG (yttrium aluminum garnet) laser with wavelength of 1064 nm, repetition frequency of 20 Hz, fluence of 15mW/pulse, pass through a half-wave plate and illuminate a specimen, and SH waves with wavelength of 532 nm generated in a specimen are collected by an objective followed by an infra-red absorption filter, a multi-layer interference filter for 532 nm and an analyzer. Two-dimensional (2D) distribution of SH intensity inside a specimen is obtained by a charge coupled device (CCD) camera with image-intensifier, which is connected to a computer.

The specimen is put in a liquid He cryostat for microscopy (CF2101, Oxford instruments) and the temperature of the specimen is cooled down to 23K with a speed of 3K/min, while 1K/min on heating. Shutter speed (from 5 to 15 sec) and gain of CCD camera are selected depending on the SH intensity. Then the temperature change during the exposure time is between 0.1 and 0.75 K.

When the history dependence of SH intensity is observed, the polarizer and the analyzer are fixed along the [001] direction parallel to the applied electric field, while they are rotated along the optic axis when the anisotropy of SH intensity is observed. Comparatively homogeneous areas of 0.075 x 0.075 mm$^2$ without a scratch are selected in the 2-dimensional SH images and the average intensity are obtained by summing up photon numbers counted by a photo-detector array.

B. X-ray diffraction measurements

A specimen for the X-ray diffraction measurements was prepared by crushing a single crystal of KTL:2.8%Li with small force as possible. The sample powder is mounted on a gilded Cu plate with a small amount of silicon grease. The measurement is performed using a horizontal-type goniometer for powder samples (Rigaku, RINT-TTR3C) equipped with a He cryostat (Rigaku 4K1) on heating from 10 K with heating speed of 0.5 K/min. Rotating Cu anode ($\lambda_{K\alpha}$=1.5418A) is used under the condition of 50kV and 300mA. The 400 reflection with 2θ=100~103 degree is used to determine the lattice constants using a bended graphite monochromator, a divergence slit of 1/2 degree and a receiving slit of 0.15 mm.

## III. EXPERIMENTAL RESULTS

A. Observation of the history dependence of SH intensity

The specimen is cooled down to 24K, then the SH image is observed on heating the specimen up to 100K. No marked SH intensity is observed in this process (ZFH/ZFC). Then the specimen is cooled down again down to 24.8K. The electric field $E$ is applied at the temperature and the specimen is heated up to 60K under the same field strength. In this process (FH/ZFC), weak SH intensity observed at low temperature starts to increase abruptly at certain temperature $T_1$, shows a peak and decreases and finally vanishes at a temperature $T_2$, as shown in Fig. 3(a). $T_1$ and $T_2$ depend on the field strength in Fig.4: The stronger is the field, the lower $T_1$, the higher $T_2$, and the stronger the SH intensity. It should be also noted that the SHG intensity is not homogeneously distributed in the sample as shown in SH images (Fig.3), where right spots and dark regions coexist. As the average diameter of bright spots is 1μm, these are not nano-polar clusters themselves, but the accumulated clusters or the scattered SHG by polar nano-clusters. The dark regions correspond to regions where macroscopic polarization does not develop. Although we cannot perfectly eliminate the possibility that SH intensity vanishes as a result of random distribution of the phase of SH waves produced by polar regions, the possibility would be small as the measurement is performed under the electric field. However, near $T_F$, the difference of the Li concentration changes the characteristic temperatures, and some parts remains bright but other parts dark. In fact, we observed the $T_F$ differs from place to place in the sample, by 4.0K. This enables us to estimate the distribution of the Li concentration to be 0.33% in the sample. It is emphasized that the SHG microscopic observations enable us to provide information on the inhomogeneity of materials, while usual SHG measurements cannot do it. In the experiment, we find that the specimen memorizes the previous experience of the field. To erase the memory effect, the specimen is kept at room temperature during 12 hours for each run.

We also observed the SH images in FH after FC process. The electric field of 80V/m is applied to the specimen at room temperature, the specimen is cooled down to 23K, then heated up with same field strength. The strong SH intensity is observed at low temperature and gradually decreases with the increase of temperature, and vanishes at 50K, as shown in Fig.3(b). A hump of the SHG intensity around 45K would be due to the imperfect erasing of the memory effect. The ZFH after FC process also takes almost same path. The vanishing temperature coincides with that observed in the FH/ZFC with same field strength. Then the temperature (=50 K) can be defined as the freezing temperature $T_F$ of KTL:2.8%Li. Temperature dependences of SH intensity of different paths are summarized in Fig5.

B. Anisotropy of SH intensity at low temperature

In order to determine the average symmetry of the field-induced polar state of KTL:2.8%Li, polarization dependences of SH intensity were measured at 23K after FC process. The procedure of determining the SH intensity is same as in the measurement of the history dependence.

Fig.6(a) shows the SHG intensity as a function of the rotation angle of the polarizer measured with the analyzer fixed parallel to the direction of the electric field (//[001]) (Case I). A sinusoidal curve of the periodicity of 180 degree is observed: The maximum is obtained when the polarization direction of the fundamental wave is parallel to the electric field. Similar measurement is made with the analyzer fixed perpendicular to the field direction (Case II). A sinusoidal curve of the periodicity of 90 degree is obtained as shown in Fig.6(b). Minima are obtained when the polarization direction of the fundamental wave is parallel and perpendicular to the field direction. Then the SH intensities are measured as a function of the polarization direction of the SH wave with the polarizer fixed parallel (Case III) and perpendicular to the field direction (Case IV), and results are shown in Fig.6(c) and (d), respectively. These results are best fitted with theoretical curves assuming the point group of tetragonal 4mm as discussed in the session IV.

C. Temperature dependence of lattice constants

Powder diffraction profile of (400) reflection spectrum was measured from 10K with a step of 10K up to 100K in ZFH/ZFC process. Examples of the result are indicated in Fig.7. At low temperature, a clear split of the profile is observed as shown in Fig.7(a) and (b). On the other hand, above $T_F$, a split disappears and only a well defined peak is observed (Fig.7(c)). Lattice constants determined from the analysis are plotted as a function of temperature in Fig.8. Results of KTO, KTL:1.6%Li and 5%Li determined by Andrews[11] are also shown for comparison. Our data are located just between those of KTL:1.6%Li and KTL:5%Li. So it is naturally concluded that the macroscopic symmetry of KTL:2.8%Li is tetragonal.

## IV. DISCUSSIONS

A. A picture of the polar state of KTL:2.8%Li

Existence of the difference between ZFC and FC has already been reported implicitly, but the present paper discloses the history dependence in more systematic manner. In particular, the peculiar behavior of FH/ZFC is reported for the first time. This indicates a strong evidence for the fact that KTL below the critical concentration is disordered polar state, and no long-range order develops. The dipolar glass state or polar nano-clusters freezes gradually with deceasing temperature below $T_F$. The application of the external electric field aligns the direction of dipoles produced by off-center Li approximately parallel to the field direction, but it is localized only around Li impurities. At sufficiently low temperature, the freezing of the dipole is so strong that the application of the electric field cannot align the dipole at all. However when the electric field is applied above $T_F$, a number of dipoles are oriented to the field direction, and the FC process keeps the state down to low temperature. Under strong field, induced dipoles interact each other and ferroelectric-like state is realized. This is a qualitative explanation of the present experiment. It should be stressed that the history-dependence shown in Fig.5 is quite similar to that observed in relaxor $Pb(Mg_{1/3}Nb_{2/3})_3$. [32]

The fact indicates that the model of polar nano-cluster formation in cubic matrix is more relevant in KTL with the critical concentration. The present study did not determine nano-cluster size or correlation length of dipole glass. However several studies have already tackled the problem and obtained a consistent result of the size around 20nm. [22]

B. Average symmetry of the polar state induced by electric field

To explain the results of Fig.6, we first assume the point group tetragonal 4mm with the polar axis parallel to the cubic [001]. Non-zero SHG tensor components $d_{ij}$ (the Voigt notation is adopted here. i=1~3, j=1~6) are described as follows.

$$\begin{pmatrix} \cdot & \cdot & \cdot & \cdot & d_{15} & \cdot \\ \cdot & \cdot & \cdot & d_{15} & \cdot & \cdot \\ d_{15} & d_{15} & d_{33} & \cdot & \cdot & \cdot \end{pmatrix}$$

Here Kleiman's law on a transparent nonlinear optical crystal is taken into account. There exist two independent components $d_{15}$ and $d_{33}$.

SHG intensities $I_1$, $I_2$, $I_3$ and $I_4$ corresponding to Cases (I), (II), (III) and (IV) are expressed as follows.

$$I_1 = I_{10}[\sin^2(\theta - \theta_{10}) + p_1 \cos^2(\theta - \theta_{10})]^2 \tag{1}$$

$$I_2 = I_{20}[p_2 \cos(\theta - \theta_{20}) \sin(\theta - \theta_{20})]^2 \tag{2}$$

$$I_3 = I_{30}[p_3 \cos(\theta - \theta_{30})]^2 \tag{3}$$

$$I_4 = I_{40}[p_4 \cos(\theta - \theta_{40})]^2 \tag{4}$$

Here $I_{i0}$ (i = 1~4) means the intensity of fundamental laser wave, $\theta_{i0}$ the direction of the electric field (//[001]), and $p_i$ is expressed using SHG tensor components as

$$p_1 = d_{33}/d_{32} = d_{33}/d_{15}, \quad p_2 = 2d_{15}, \quad p_3 = d_{33}, \quad p_4 = d_{32} = d_{15} \tag{5}$$

Values determined by fitting experimental results in Fig.6 are

$$p_1 = 2.82, \quad p_2 = 1.87, \quad p_3 = 2.63, \quad p_4 = 0.92 \tag{6}$$

From these values, the following constant (in arbitrary unit) are consistently obtained.

$$d_{33}/d_{15} = 2.86 \pm 0.18 \tag{7}$$

These values are almost same as the previous reports as shown in Table I.

We assumed that the point group is tetragonal 4mm, but our results can be also explained by tetragonal 4, or lower symmetry point group, e.g., orthorhombic mm2. However the result of X-ray diffraction experiments shows that the most plausible crystal system is tetragonal. Then the present experiment concludes that the average symmetry of field-induced phase of KTL with Li critical concentration is 4mm or 4.

## V. CONCLUSIONS

Under the SHG microscope, we observed marked history dependence of the polar phase induced by the electric field in KTL with critical concentration of Li (2.8 mol%). ZFH/ZFC, FH/ZFC, FH/FC or ZFH/FC processes follow different paths. The result indicates that the ground state of KTL with the critical concentration of Li is intrinsically inhomogeneous where the polar state is localized and the long-range ferroelectric state is not developed. Two models would be plausible: one is polar nano-cluster model similar to relaxors, and the other is a dipole glass model. The formation of polar nano-clusters has already reported in pure KTO by using Raman scattering measurements.[33]

The history dependence of the order parameter similar to prototype relaxor PMN seems to support the polar nano-cluster model. The marked photo-current observed in low temperature region is also consistent with the model, by considering the percorative nature of nano-clusters: If the nano-cluster has semiconductive nature, they are connected under the field to generate conductive nets on illumination. For understanding the phenomenon more quantitatively, further experiments are necessary, especially that for disclosing semiconductive nature of the polar nano-cluster, e.g. by using scanning prove microscopes. The X-ray diffraction experiments using a single crystal is also waited for.

The work is supported by the grants-in-aid of scientific research (A) of MEXT, the special research program of Waseda University and the 21st century COE program "Physics of systems with self-organization composed of multi-elements" of MEXT,

Japan. We are grateful to Prof.Y.Yamada, Prof.Y.Tsunoda, Prof.K.Kohn, Waseda Univ., Dr.J.M.Kiat, Ecole Centrale Paris, for valuable discussions, and Mr.Komaki for his help to X-ray measurements, and Mr.Kinoshita for growing single crystals.

Table I. Experimentally determined SHG tensor component ratio
$d_{33}/d_{15}$ of KTL:$x$%Li

| $d_{33}/d_{15}$ | Li concentration $x$% | E(V/mm) | Reference |
|---|---|---|---|
| 2.86±0.18 | 2.8 | 80 | Present work |
| 2.51±0.12 | 0 | 200 | 19 |
| 2.51±0.12 | 3.6 | 0, 80 | 19 |
| 2.45±0.2 | 1.6, 2.6, 3.4, 6 | 0 | 22 |
| 2.70±4% | 0 | 2050 | 33 |
| 3.03±4% | 0 | 1530 | 33 |

Figure caption

Fig.1 The orientation and dimension of the specimen used for SHG microscopic observations. A pair of Au electrodes are evaporated on the top surface of KTL. The laser illuminated area is located between two electrodes to avoid the photo-current effect. The direction of electric field is parallel to [001] referred to the cubic axes.

Fig.2 Optical system of the SHG microscope. (1) CCD camera with image intensifier, (2) 532nm-pass filter, (3) and (5) infrared absorption filter, (4) objective lens, (6) He cryostat, (7) half wave plate, (8) reflective mirror.

Fig.3 SHG images of KTL:2.8%Li. (a) indicates those observed in the FH after ZFC process, (b) in the FH after FC. Brighter parts produce stronger SH waves. In (a), SHG appears only in a limited temperature range below the freezing temperature $T_F$ of 50 K, while in (b), SHG is observed in whole temperature range below $T_F$.

Fig.4 Temperature dependences of SH intensities of KTL:2.8%Li measured in the FH after ZFC process under different field strengths. In each process, the specimen was kept in room temperature for 12 hours before starting the measurement.

Fig.5 Temperature dependences of SH intensity of KTL:2.8%Li in ZFC, FH after ZFC and FH after FC processes. A hump observed near 45K in FH after FC process could be due to an imperfect erasing of the memory effect of the previous measurement.

Fig.6 Polarization dependences of SH intensity of KTL:2.8%. (a) indicates the case (I) of rotating the polarizer with the analyzer fixed parallel to the electric field E(//[001]), (b) the case (II) of rotating the polarizer with the analyzer fixed perpendicular to E, (c) the case (III) of rotating the analyzer with polarizer fixed

parallel to E, and (d) the case (IV) of rotating the analyzer with polarizer fixed perpendicular to E. Solid lines indicate fitted curved calculated using Eqs.(1) ~ (4).

Fig.7　X-ray powder diffraction profiles of KTL:2.8%. (a) indicates the profile at T= 10K, (b) at T= 30K, and (c) T = 70K., where the Lorentz function is used for the analysis.

Fig.8　Temperature dependences of lattice constants of KTL:2.8%. Results of *x*=0, 1.6 and 5 determined by Andrews[10] are also plotted for comparison.

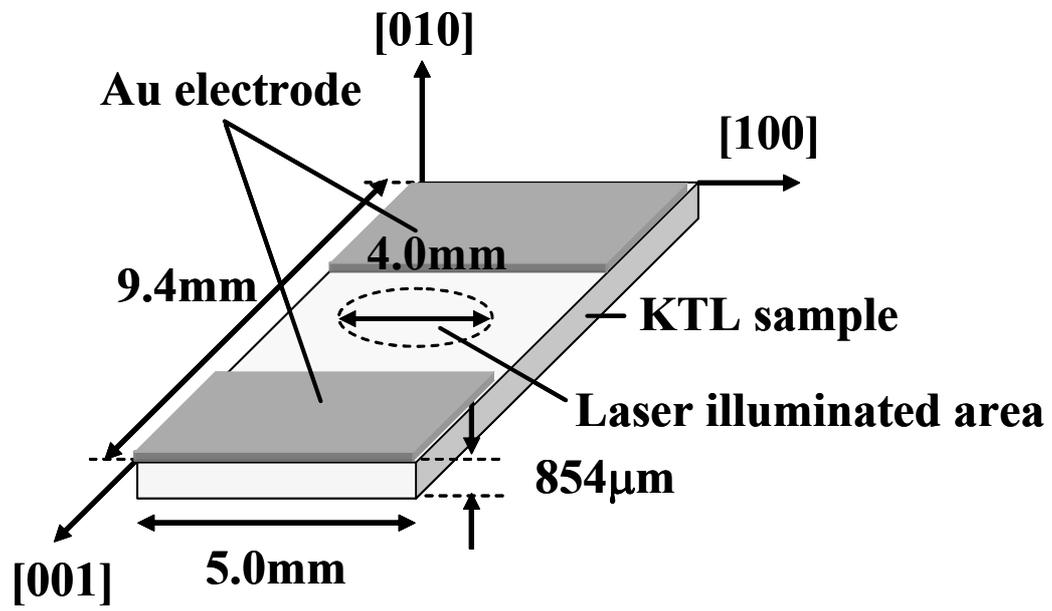

FIG. 1. H.Yokota et al, PRB.

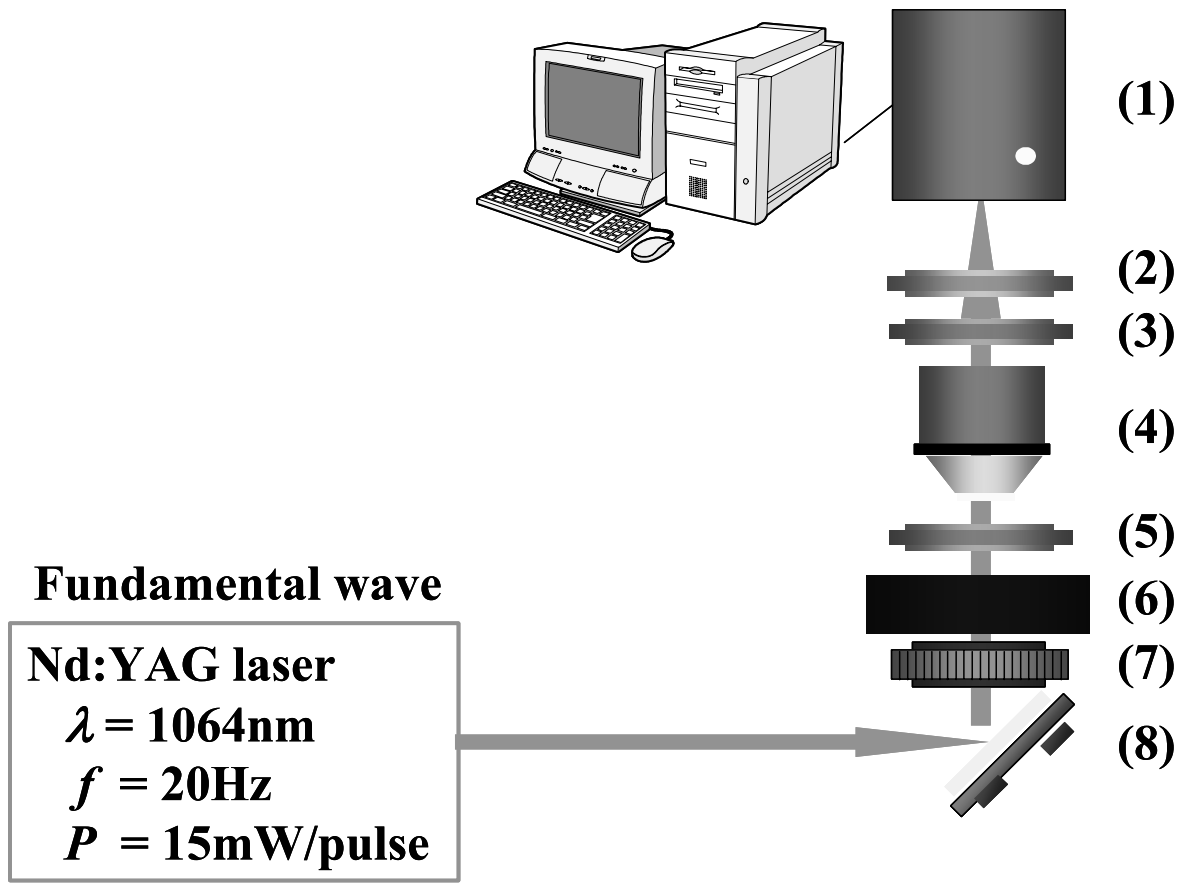

FIG. 2. H.Yokota et al, PRB

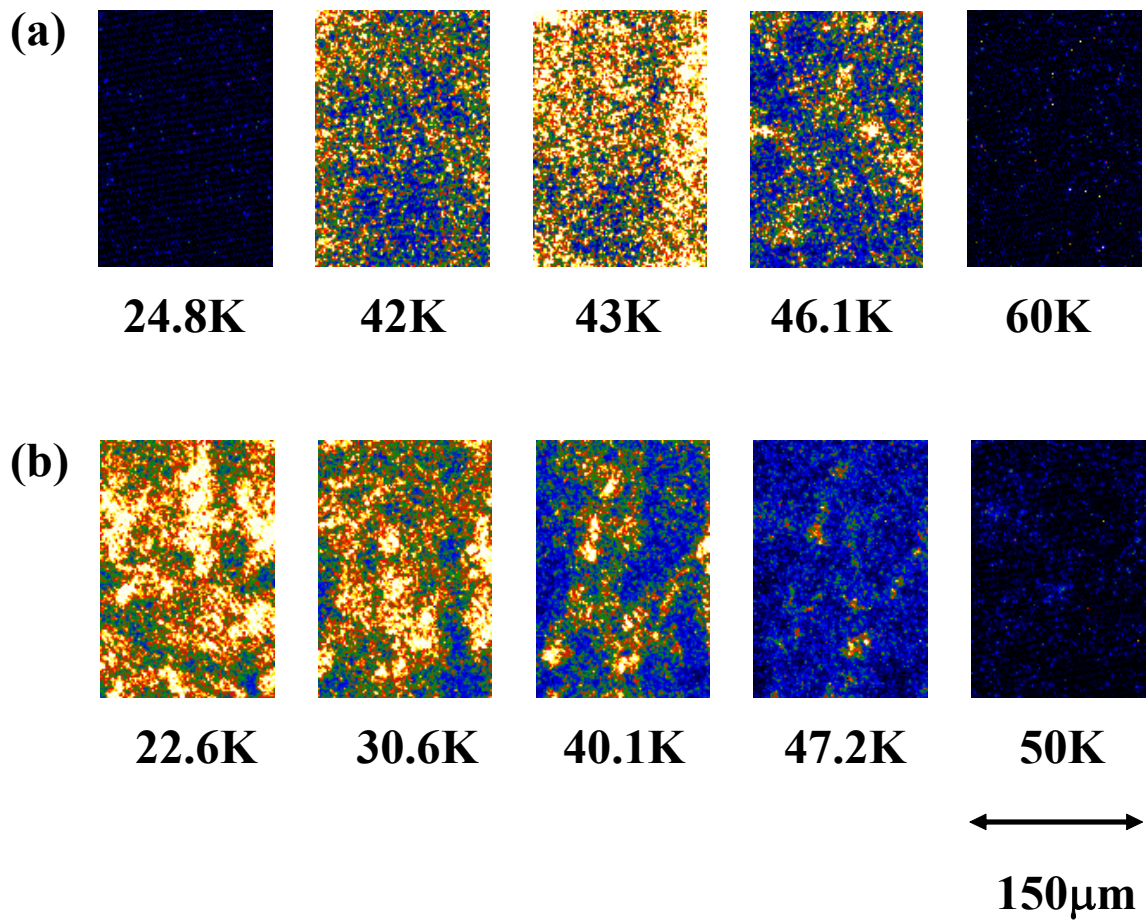

FIG. 3. H.Yokota et al, PRB

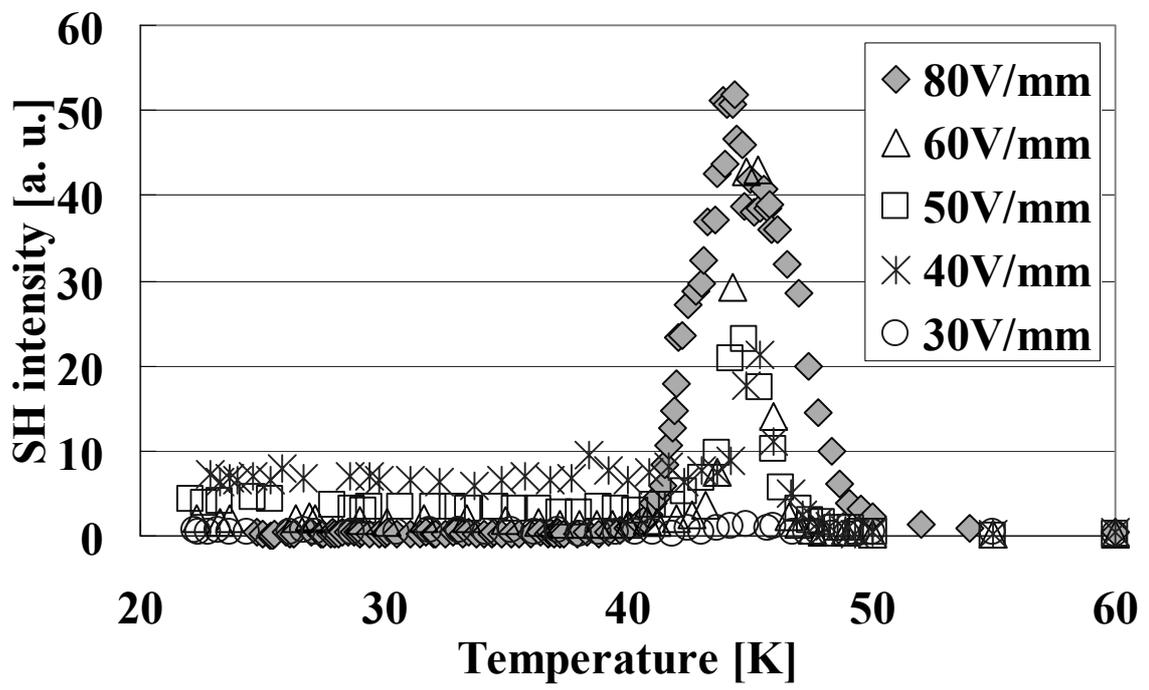

FIG. 4. H.Yokota et al, PRB

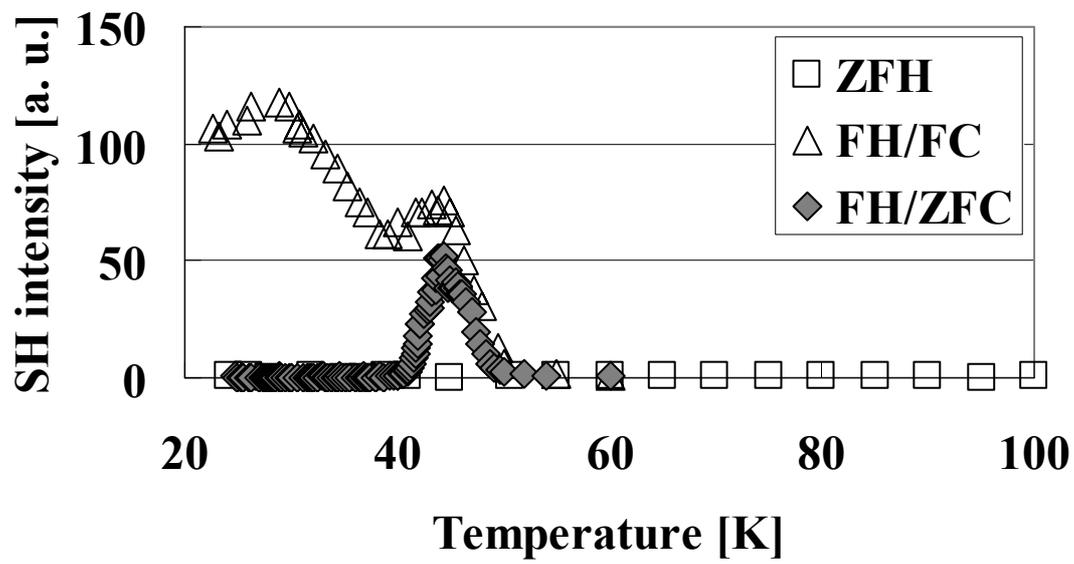

FIG. 5. H.Yokota et al, PRB

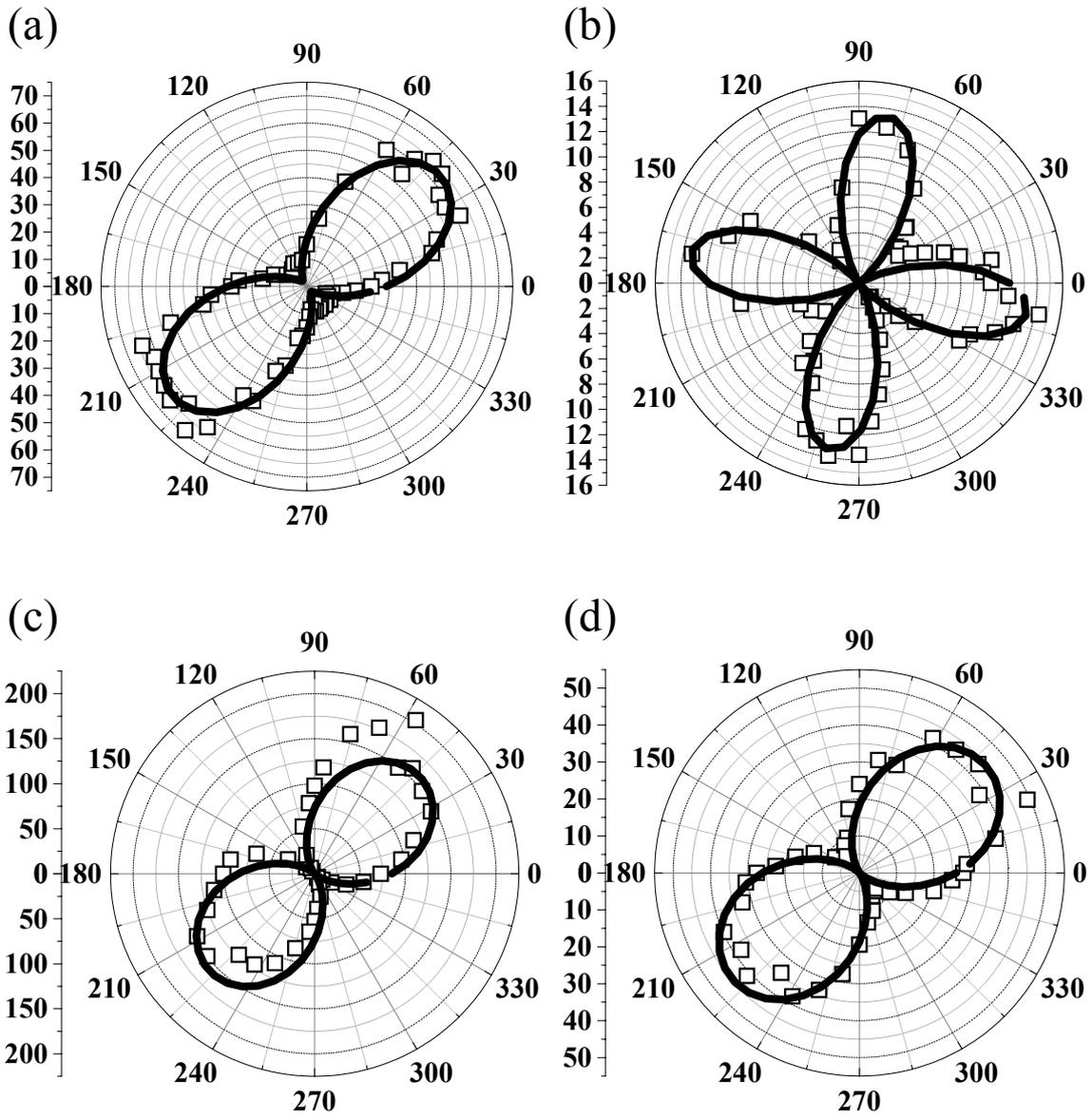

FIG. 6. H. Yokota et al, PRB

(a)

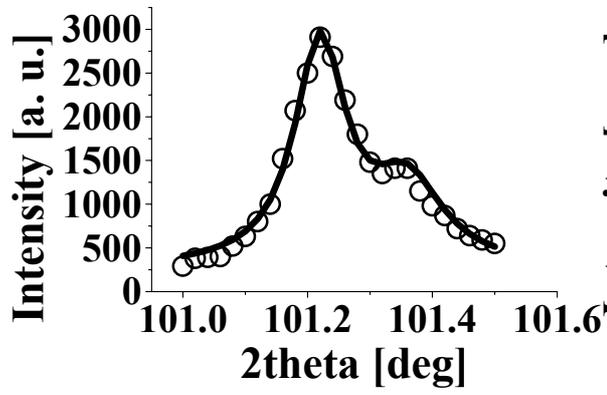

(b)

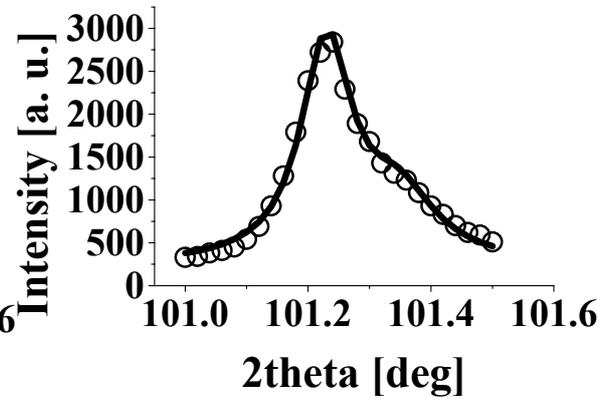

(c)

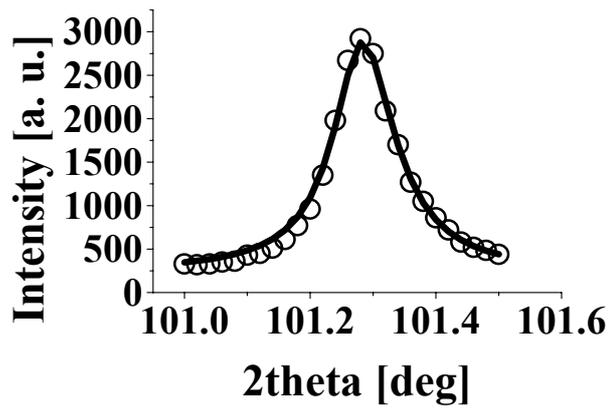

FIG. 7. H. Yokota et al, PRB

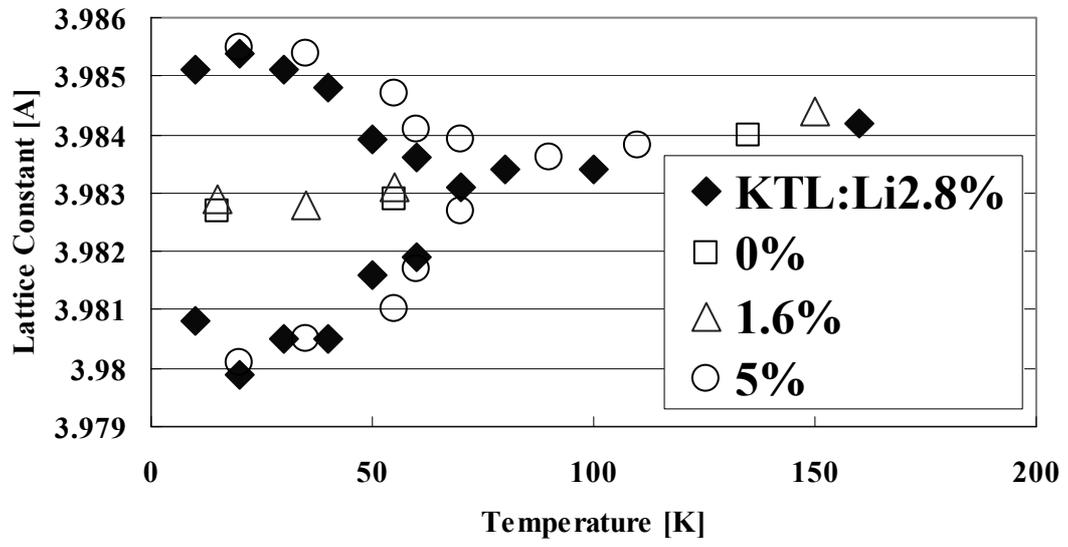

FIG. 8. H. Yokota et al, PRB